\documentclass{ws-procs9x6}

\begin{document}

\title{
Contrasting Pathways to Mott Gap Collapse in Electron and Hole Doped Cuprates}

\author{R. S. Markiewicz}

\address{Northeastern University, 360 Huntington Avenue,  
Boston MA 02115, USA\\ 
E-mail: markiewic@neu.edu}

\maketitle

\abstracts{
Recent ARPES measurements on the electron-doped cuprate Nd$_{2-x}$Ce$_x$CuO$_4$
can be interpreted in a mean field model of uniform doping of an
antiferromagnet, with the Mott gap closing near optimal doping.
Mode coupling calculations confirm the mean field results, while
clarifying the relation between the Mott gap and short-range magnetic order.
The {\it same} calculations find that hole doped cuprates should follow a
strikingly different doping dependence, involving instability toward spiral
phases or stripes.  Nevertheless, the magnetic order (now associated with
stripes) again collapses near optimal doping.}

\section{Introduction}
\subsection{Mode Coupling Theories}\label{subsec:prod}
In conventional itinerant ferro- and antiferromagnets, mode-coupling 
theories\cite{MuDo,Mor} have proven of value in treating the role of 
fluctuations in reducing or eliminating long-range order, as well as in the 
development of local moments, Curie-like susceptibility, and in general the 
crossover to magnetic insulators.  Similar approaches have been applied to 
charge-density wave systems\cite{SuMo} and the glass transition\cite{Got}.  

Attempts to apply such a formalism to study the antiferromagnetism of
the cuprate superconducting compounds have been frustrated, since the 
antiferromagnetic (AF) phase is found to be unstable against hole doping, toward
either an incommensurate AF phase\cite{ShSi} or phase separation\cite{Schulz}. 
Here, it is 
demonstrated that this tendency to instability is either absent or greatly
reduced in electron-doped cuprates, and the mode coupling analysis can provide
a detailed description of the collapse of the Mott gap with doping.  The results
are of great interest of themselves: the collapse is associated with one or more
quantum critical points (QCPs), and superconductivity is optimized close to one
QCP.  However, the results have an additional importance in the light they
shed on the more complicated problem of the {\it hole doped} cuprates.  First,
the effective Hubbard $U$ parameter has a significant doping dependence, from
$U\sim W$ at half filling (where $W=8t$ is the bandwidth and $t$ the nearest
neighbor hopping parameter) to $U\sim W/2$ near optimal doping.  Secondly, 
the {\it same} mode coupling theory which works so well for electron-doped
cuprates, {\it breaks down} for hole doping, due to the above noted electronic
instability.  This suggests that (a) the electron-hole asymmetry must be due to
a band structure effect and (b) a suitable generalization of mode coupling
theory should be able to incorporate the effects of this instability.

\subsection{ARPES of Nd$_{2-x}$Ce$_x$CuO$_4$}
Recently, Armitage, et al.\cite{nparm} succeeded in measuring angle-resolved
photoemission spectra (ARPES) of Nd$_{2-x}$Ce$_x$CuO$_4$ (NCCO) as a function
of electron doping, from essentially the undoped insulator $x=0$ to optimal 
doping $x=-0.15$.  The doping dependence is strikingly different from that
found in  {\it hole-doped} La$_{2-x}$Sr$_x$CuO$_4$ (LSCO)\cite{ZZX,Ino}. 
Both systems start from a Mott insulator at half filling, with ARPES being
sensitive only to the lower Hubbard band (LHB), approximately 1eV below the
Fermi level $E_F$.  With hole doping, the LHB remains well below $E_F$ while
holes are added in mid-gap states (as expected, e.g., in the presence of
nanoscale phase separation\cite{tfstr}); for electron doping, the Fermi level 
shifts to the upper Hubbard band (UHB), and the electrons appear to uniformly
dope the antiferromagnet.

Remarkably, the full doping dependence can be simply described by a mean-field
(MF) $t - t' - U$ Hubbard model\cite{KLBM}, where $t'$ is the second neighbor
hopping, $U$ is the onsite coulomb repulsion, and a one band (copper only)
model was assumed for simplicity.\footnote{Actually, in Ref.~\cite{KLBM} a
third neighbor hopping $t''$ was included to optimize the fit to the 
experimental Fermi surface curvature.  The changes induced by this parameter are
small, and it will be ignored in the present calculations.}
A key finding is that the Hubbard $U$ is doping dependent, leading to a quantum
critical point (QCP) just beyond optimal doping, where the Mott gap vanishes.
The appearance of a peak in superconductivity near an AFM QCP is a fairly
common occurance\cite{MaLo}; in particular, something similar has been 
observed\cite{Tal1} in the hole-doped cuprates.  However, the MF theory is
problematic, in that the Mott gap is associated with long-range N\'eel order,
and the MF model predicts anomalously high values for $T_N$.

\section{Mode Coupling Theory}

This anomalous behavior can be cured by incorporating the role of fluctuations.
Indeed, it is known that in a two-dimensional system, the N\'eel transition
can only occur at $T=0$ -- the Mermin-Wagner (MW) theorem\cite{mw}.   By
treating fluctuations within a mode coupling analysis, the MW theorem is
satisfied\cite{RM1}, and the Mott gap is completely decoupled from long-range
spin-density wave (SDW) order.  Even though $T_N=0$ a large Mott (pseudo)gap
is present even well above room temperature near half filling -- due
to {\it short-range} AFM order.  The MF gap and transition temperature are
found to be approximately the pseudogap and $T^*$ the onset temperature for
the pseudogap opening.

The calculation can be summarized as follows.  In a path integral formulation
of the Hubbard model\cite{NNag}, the quartic term is decoupled via a 
Hubbard-Stratonovich transformation and the fermoinic degrees integrated out.
The resulting action is then expanded to quartic order in the 
Hubbard-Stratonovich fields $\phi$.  The quadratic interaction reproduces the
RPA theory of the Hubbard model -- the quadratic coefficient is just $U\delta_{
0q}$, where $\delta_{0q}$ is the (inverse) Stoner factor $\delta_q=1-U\chi_0(
\vec Q+\vec q,\omega )$ (here $\vec Q=(\pi ,\pi )$ is the wave vector associated
with the commensurate SDW).  The quartic interaction, parametrized by the 
coefficient $u$ evaluated at $\vec Q$, $\omega =0$, represents coupling between
different magnetic modes.  The effects of this term cannot be treated in
perturbation theory, and a self-consistent renormalization (SCR) 
scheme\cite{Mor} is introduced to calculate the renormalized Stoner factor 
$\delta_q=\delta_{0q}+\lambda$.  A self-consistent equation for $\delta$ is
found, which can be solved if the band parameters and the interactions
$U$ and $u$ are known.

To simplify the calculation, $\lambda$ is assumed to be independent of $\vec q$
and $\omega$, and the Stoner factor is expanded near $\vec Q$ as 
\begin{equation}
\delta_q(\omega )=\delta +Aq^2-B\omega^2-iC\omega,
\label{eq:1}
\end{equation}
similar to the form assumed in nearly antiferromagnetic Fermi liquid 
(NAFL)\cite{NAFL} theory.  The imaginary term linear in frequency is due to
the presence of low energy magnon excitations in the vicinity of the `hot
spots' -- the points where the Fermi surface intersects the Brillouin zone 
diagonal.  Here fluctuations toward long-range N\'eel order lead to strong,
Bragg-like scattering which ultimately leads to the magnetic Brillouin zone
boundary at $T_N$.  
Independent of the parameter values, it is found that
$\delta >0$ for $T>0$ -- the MW theorem is satisfied, while for electron
doping $\delta\rightarrow 0$ as $T\rightarrow 0$ up to a critical doping -- 
there is a QCP associated with $T=0$ SDW order.  For hole doping the
calculation breaks down -- the parameter $A$ is negative in a significant
doping regime, as will be discussed further below.

The key insights of mode coupling theory are:

$\bullet$ The Mott transition is dominated by hot spot physics, which creates 
zone-edge magnons.  The condensation of these magnons creates a new zone 
boundary, and opens up a gap (the Mott gap) in the electronic spectrum.  In 
two dimensions (2D), there can be no Bose condensation at finite temperatures, 
but the pileup of lowest energy magnons as $T$ decreases leads to the appearence
of a Mott pseudogap and a T=0 transition to long-range SDW order.

$\bullet$ Evidence for the existence of local magnons comes from well-defined 
{\it plateaus} in the spin susceptibility.  Plateaus are seen in (a) the
doping dependence of the susceptibility at $\vec Q$, (b) the $\vec q$ 
dependence of the susceptibility near  $\vec Q$, and (c) the $\omega$
dependence of the susceptibility (both real and imaginary parts) at $\vec Q$.
These plateaus introduce {\it cutoffs} in the $\vec q$ and $\omega$ dependence
of the Stoner factor, Eq.~\ref{eq:1}, which in general {\it cannot} be sent to
$\infty$, in contrast to the NAFL model.  Also, the flatness of the plateau tops
makes it difficult to estimate the model parameters from first principles.  In 
particular $A$ is strongly temperature dependent, while $u$ is anomalously
small (this problem had been noted previously\cite{Chu}).

$\bullet$ A finite N\'eel temperature can be generated by interlayer coupling.  
In the cuprates, such coupling is typically frustrated, and N\'eel order more 
likely arises from a Kosterlitz-Thouless transition, after the spin 
dimensionality is reduced by, e.g., spin-orbit coupling\cite{Ding}.

\section{Results}
\subsection{Susceptibility}

In analyzing the ARPES data, a tight-binding band is assumed, 
\begin{equation}
\epsilon_k=-2t(c_x+c_y)-4t'c_xc_y,
\label{eq:0}
\end{equation}
with $c_i=\cos{k_ia}$, $t=0.326eV$, and $t'/t=-0.276$.  The Hubbard $U$ is 
doping dependent, $U/t$ = 6, 5, 3.5, and 2.9 at $x$ = 0, -0.04, -0.10, and
-0.15, respectively, and the mode coupling constant is adjusted to reproduce
the low-temperature spin stiffness at half filling\cite{CHN}, $u^{-1}=0.256eV$.
The bare susceptibility
\begin{equation}
\chi_0(\vec q,\omega) =-\sum_{\vec k}{f(\epsilon_{\vec k})-f(\epsilon_{\vec k+
\vec q})\over\epsilon_{\vec k}-\epsilon_{\vec k+\vec q}+\omega+i\delta},
\label{eq:2}
\end{equation}
is evaluated near $\vec Q$ to determine the parameters of Eq.~\ref{eq:1}.

\begin{figure}[t]
\epsfxsize=20pc 
\epsfbox{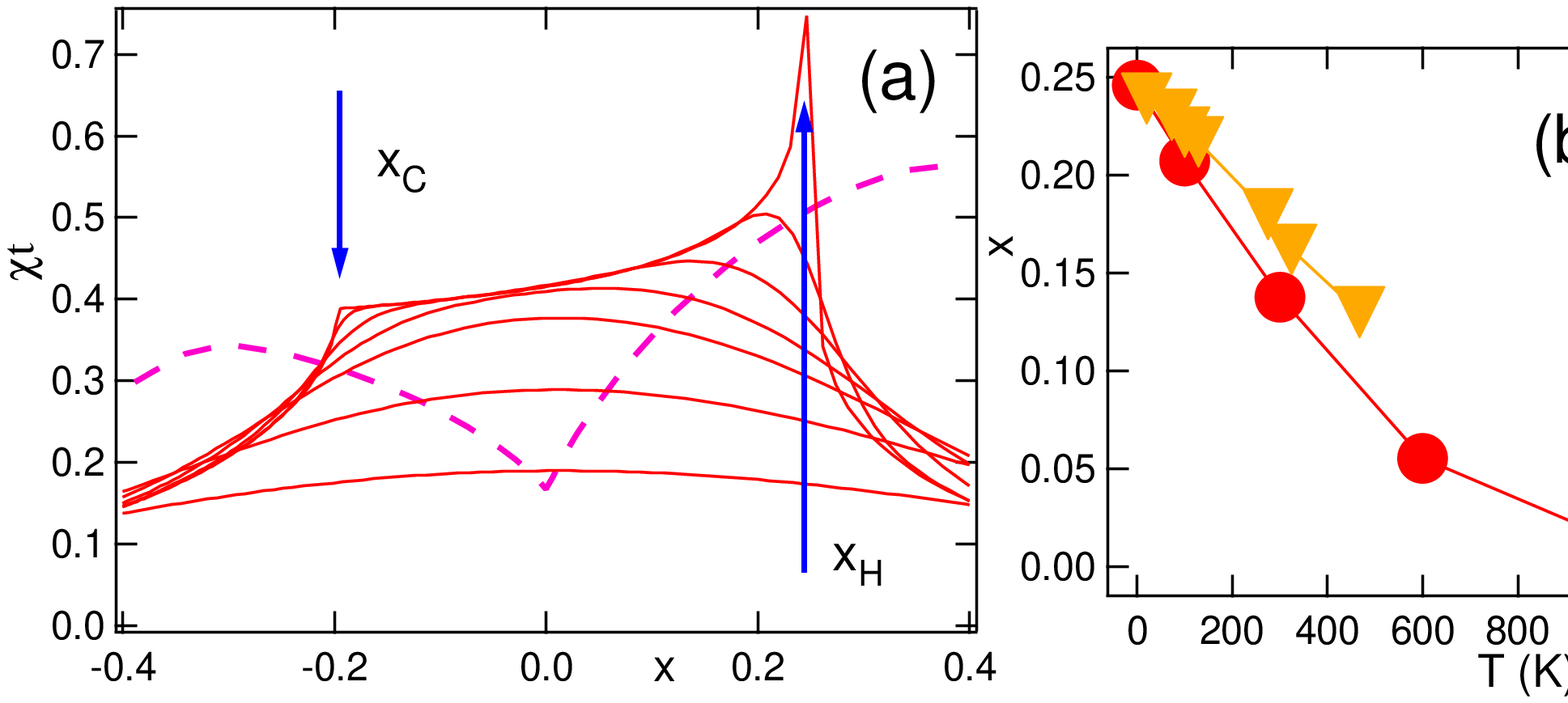} 
\caption{(a) Susceptibility $\chi_0$ at $\vec Q$ as a function of doping for
several temperatures.  From highest to lowest curves near $x=0.1$, the
temperatures are $T$ = 1, 100, 300, 600, 1000, 2000, and 4000 K. Dashed line =
$1/U_{eff}$.  (b) Circles = pseudo-VHS (peak of $\chi_0$) as a function of
temperature $T_V$; triangles = $T_{incomm}$.
\label{fig:1}}
\end{figure}
\begin{figure}
\epsfxsize=20pc 
\epsfbox{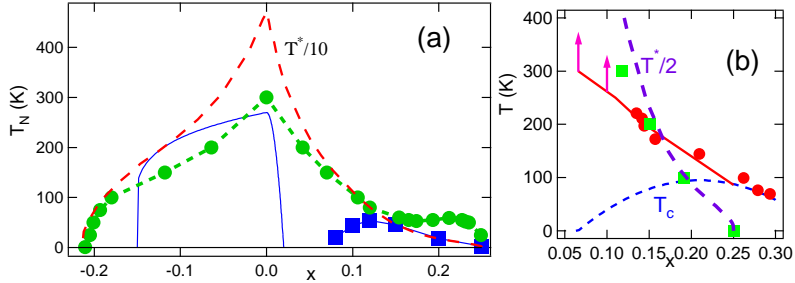} 
\caption{(a) Mean-field transition temperature $T^*$ (long dashed line) 
compared with N\'eel temperature of NCCO and LSCO (solid line) plus magnetic
transition temperature of stripe ordered phase in Nd-substituted 
LSCO\protect\cite{Ich},  Circles = model calculation of $T_N$ assuming
unfrustrated interlayer hopping. (b) Comparison of mean-field transition 
(long dashed line) to various estimates of pseudogap temperature: from
photoemission\protect\cite{Camp} (solid line), heat capacity\protect\cite{Tal0} 
(squares) and tunneling\protect\cite{Miya} (circles = $\Delta /3$, with $\Delta$
the tunneling gap). Short dashed line = superconducting $T_c$.
\label{fig:2}}
\end{figure}
The susceptibility $\chi_0 (\vec Q,0)$ has approximately the shape of a 
plateau as a function of doping, Figure~\ref{fig:1}a, bounded by the critical 
points $x_H$ and $x_C$ where the Fermi surface ceases to have hot spots.  These
special points act as {\it natural phase boundaries} for antiferromagnetism,
due to the sharp falloff in $\chi$ off of the plateau.

The special point $x_H$ coincides at $T=0$ with the Van Hove singularity (VHS) 
of the band.  Remarkably, the susceptibility peak has a strong temperature
dependence\cite{OPfeut}, defining a pseudo-VHS; Fig.~\ref{fig:1}b shows the 
temperature $T_V(x)$, at which the susceptibility peaks at $x$.  This can be 
understood by noting that the energy denominator of $\chi$, Eq.~\ref{eq:2}, is 
{\it independent of $t'$}, and thus would lead to a large peak at half filling, 
x=0 (associated with states along the zone diagonal).
At low temperatures, the difference in Fermi functions in the numerator cuts 
this off, and forces the peak to coincide with the VHS.  As $T$ increases, 
more states near the zone diagonals become available, causing the peak
susceptibility to shift towards half filling.  For a temperature $T_{incomm}$
slightly above $T_V$, the curvature $A$ becomes negative, signalling the
instability of the commensurate AFM state.

The dashed line in Fig.~\ref{fig:1}a represents $1/U_{eff}$, where $U_{eff}$ is
a doping dependent Hubbard $U$, estimated from a screening 
calculation\cite{KLBM}.  The intersection of the dashed line with one of the
solid lines defines the mean-field N\'eel transition, $\chi_0U_{eff}=1$, 
Fig.~\ref{fig:2}a (long-dashed line).  As will be shown below, once fluctuations
are included, the mean field transition turns into a pseudogap onset $T^*$, 
while the actual onset of long range magnetic order is suppressed to much lower 
temperatures.  From Fig.~\ref{fig:2}b, it can be seen that the mean-field $T^*$
is consistent with a number of estimates\cite{Camp,Tal0,Miya} of the
experimental pseudogap for {\it hole} doping, while a simple calculation of
the three-dimensional N\'eel transition associated with interlayer coupling
(circles in Fig.~\ref{fig:2}a -- see the Appendix 
for details) can
approximately reproduce the experimental observations (solid lines) -- if
the transitions associated with magnetic order on quasistatic stripes\cite{Ich} 
are included (squares).

\begin{figure}[t]
\leavevmode
   \epsfxsize=0.75\textwidth\epsfbox{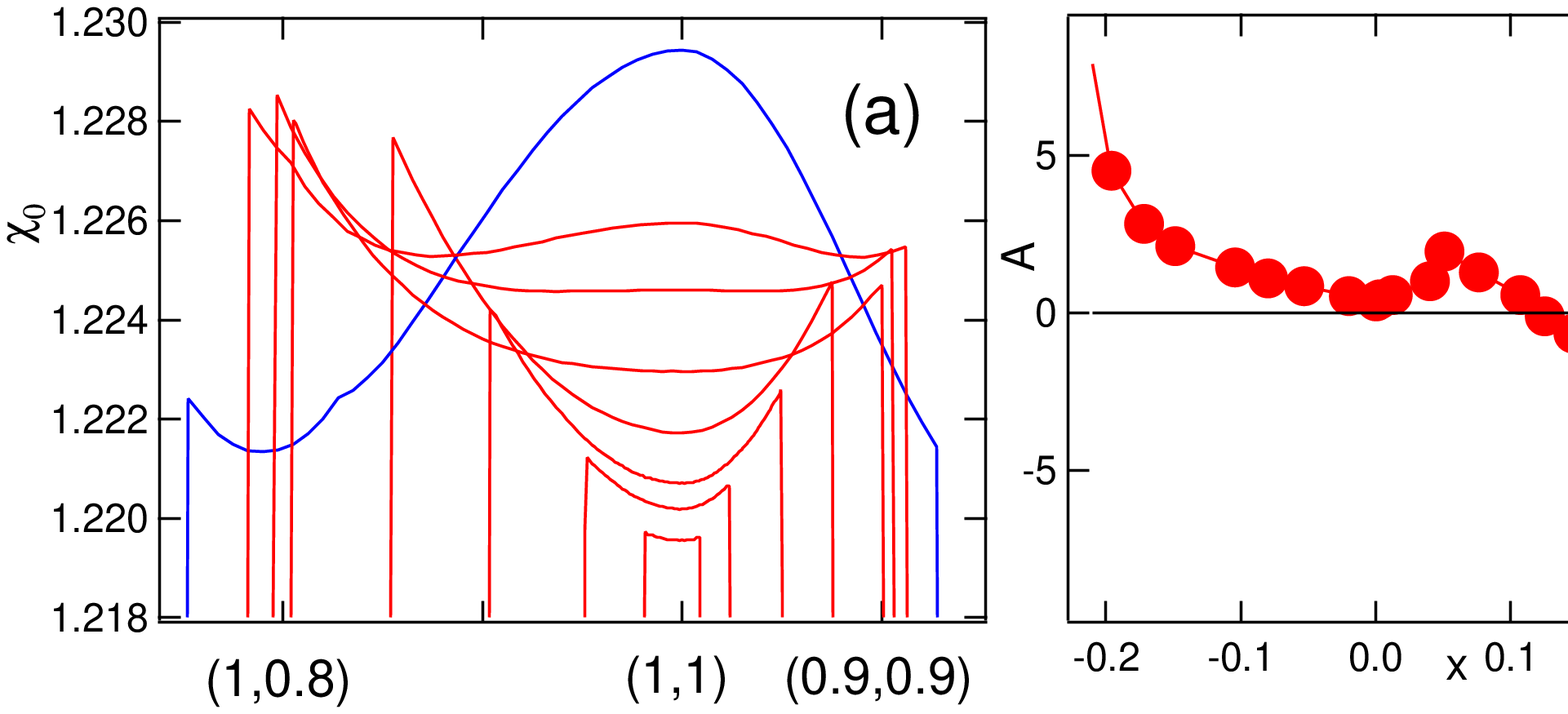}
\vskip0.5cm
\caption{(a) Expanded view of susceptibility $\chi_0$ on the plateaus near 
$\vec Q$ for a variety of dopings at $T=1K$. From highest to lowest curves near 
$\vec Q$, the chemical potentials are $\mu$ = -0.25, -0.22, -0.21, -0.20 (x=0), 
-0.15, -0.10, -0.05, and -0.02eV.  All curves except $\mu=-0.20eV$ have been 
shifted vertically to fit within the expanded frame. (b) Calculated $A(x)$.}
\label{fig:3}
\end{figure}

In addition to plateaus in {\it doping}, the hot spots lead to plateaus in 
the frequency and wave number dependence of $\chi_0(\vec q,\omega )$.  For
instance, Fig.~\ref{fig:3}a shows plateaus in $\chi_0(\vec Q+\vec q,0)$ at a
series of dopings at $T=1K$.  Once $\chi_0$ is known the parameters $A$ and $C$ 
of Eq.~\ref{eq:1} can be calculated; $B$ is quite small, and can in general be 
neglected.  The plateau width $q_c\rightarrow 0$ as $x\rightarrow x_C$, leading 
to a strong $T$-dependence of $A$, Fig.~\ref{fig:3}b.

                                                                   +

Given the model parameters, the self-consistent equation for $\delta$ 
becomes\cite{RM1}
\begin{equation}
\delta =\delta_0+
{3ua^2\over\pi^2C}\int_0^{q_c^2}dq^{'2}\int_0^{\alpha_{\omega}}dx
coth\bigl({x\over 2CT}\bigr){x\over (\delta +Aq^{'2})^2+x^2}.
\label{eq:3}
\end{equation}
which has the approximate solution
\begin{equation}
Z\delta -\bar\delta_0={3ua^2T\over\pi A}ln({2CT\over e\delta}),
\label{eq:4}
\end{equation}
where $\bar\delta_0=\delta_0+\eta -1$,
\begin{equation}
\eta =1+{3uq_c^2a^2\over \pi^2C}\bigl({1\over 2}ln[1+a_q^{-2}]+{tan^{-1}(a_q)
\over a_q}\bigr),
\label{eq:5}
\end{equation}
$a_q=Aq_c^2/\alpha_{\omega}$, and $q_c$ and $\alpha_{\omega}$ are wavenumber
and (normalized) frequency cutoffs, respectively, and the exact form of $Z$ is
not required.  From the logarithm in Eq.~\ref{eq:4}, $\delta$ must be greater
than zero for all $T>0$, so there is no finite temperature phase transition.  
At low temperatures, the $\delta$ on the left hand side of 
Eq.~\ref{eq:4} can be neglected, leading to a correlation length
\begin{equation}
\xi^2={A\over\delta}=\xi_0^2e^{4\pi\rho_s/T},
\label{eq:6}
\end{equation}
with $\xi_0^2={eA/2TC}$ and 
\begin{equation}
\rho_s={\pi A|\bar\delta_0|\over 24ua^2}.
\label{eq:7}
\end{equation}
Here, $\bar\delta_0$ is the quantum corrected Stoner factor, {\it which controls
the $T=0$ QCP}: there is long-range N\'eel order at $T=0$ whenever $\bar\delta_0
\le 0$, or $U\chi_0\ge\eta$.

\subsection{ARPES Data}

From the susceptibility, the contribution to the electronic self energy due to
one magnon scattering can be calculated.  The imaginary part of the 
susceptibility can be written
\begin{eqnarray}
Im\Sigma(\vec k,\omega )
={-g^2\chi_0\over V}\sum_{\vec q}\int_{-\alpha_{\omega}/C}^{\alpha_{\omega}/C}
d\epsilon [n(\epsilon )+f(\xi_{\vec k+\vec q})]\times
\nonumber \\
\times\delta(\omega +\epsilon -\xi_{\vec k+\vec q}){C\epsilon
\over (\delta +Aq^{'2})^2+(C\epsilon )^2}.
\label{eq:8}
\end{eqnarray}
where the coupling is approximately $g^2\chi_0\simeq{3U/2}$.  Since $\epsilon$ 
is peaked near zero when $\vec q\simeq\vec Q$, $Im\Sigma$ is approximately a
$\delta$-function at $\xi_{\vec k+\vec Q}$.  Approximating $Im\Sigma =-\pi\bar
\Delta^2\delta (\omega -\xi_{\vec k+\vec Q})$, then
\begin{equation}
\bar\Delta^2={U\over 8u}(\delta -\delta_0).
\label{eq:9}
\end{equation}

The importance of this result can be seen by noting that, by Kramers-Kronig,
\begin{equation}
Re\Sigma (\vec k,\omega )={\bar\Delta^2\over\omega -\xi_{\vec k+\vec Q}},
\label{eq:10}
\end{equation}
so that
\begin{equation}
G^{-1}(\vec k,\omega )=\omega -\xi_{\vec k}-Re\Sigma^R(\vec k,\omega )=
{(\omega -\xi_{\vec k})(\omega -\xi_{\vec k+
\vec Q})-\bar\Delta^2\over\omega -\xi_{\vec k+\vec Q}}.
\label{eq:11}
\end{equation}
The zeroes of $G^{-1}$ are identical to the mean field results for long-range
AFM order\cite{SWZ,KLBM}, except that the long-range gap $\Delta =U<m_z>$ is
replaced by the {\it short-range} gap $\bar\Delta\sim U\sqrt{<m_z^2>}$.  

\begin{figure}[t]
\epsfxsize=25pc 
\epsfbox{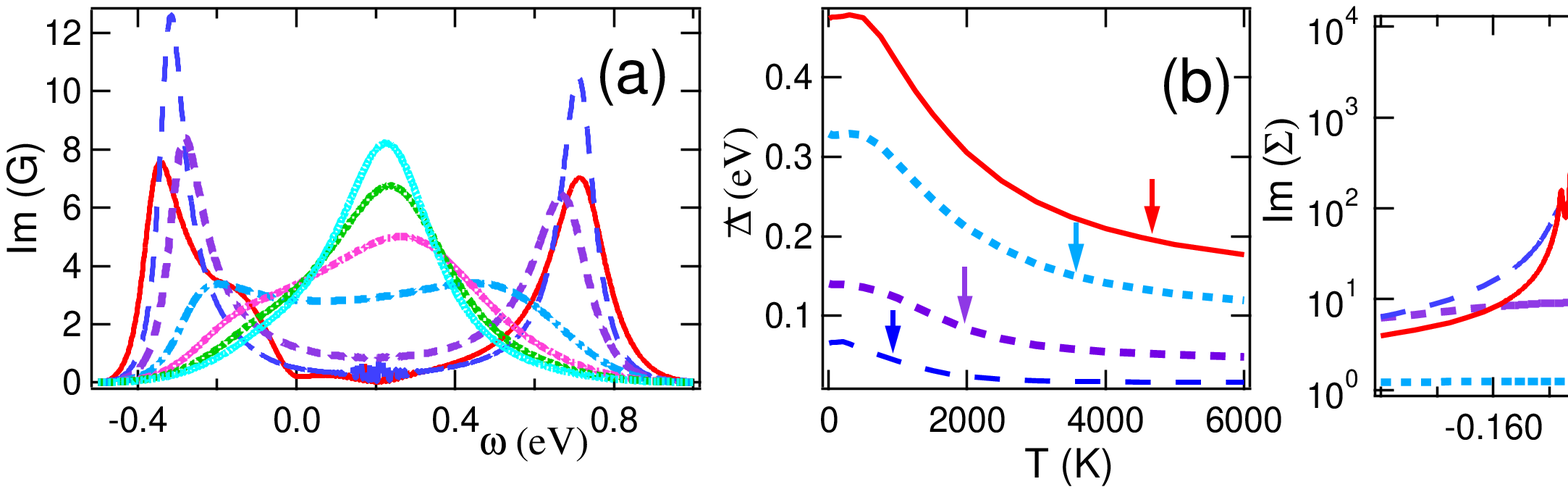} 
\caption{(a)Spectral function for $x=0.0$, $\vec k=(\pi /2,\pi /2)$, at $T$ = 
100, 500, 1000, 2000, 3000, 4000, and 5000K (larger splittings corresponding to 
lower $T$'s).
(b) $\bar\Delta (T)$; solid line: $x=0.0$, dotted lines: $x=-0.04$, short-dashed
lines: $x=-0.10$, long-dashed lines: $x=-0.15$.
(c)$Im(\Sigma )$ for $x=0$, $\vec k=(\pi ,0)$ at $T$ = 100 (solid line), 500 
(long dashed line), 1000 (short dashed line), and 2000K (dotted line).
\label{fig:5}}
\end{figure}

\begin{figure}
\epsfxsize=20pc 
\epsfbox{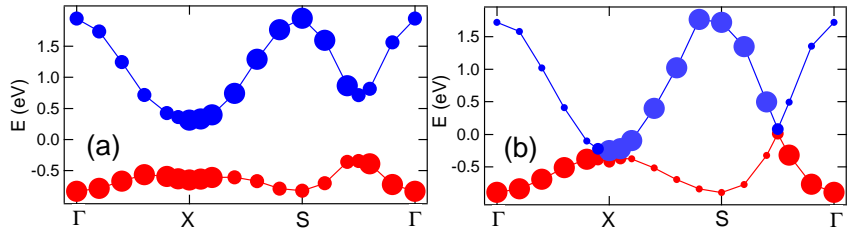} 
\caption{Dispersion relations for (a) x=0 and (b) x=-0.15, at T=100K.  Brillouin
zone directions are $\Gamma = (0,0)$, $X=(\pi ,0)$, $S=(\pi ,\pi )$.
\label{fig:6}}
\end{figure}

Figure~\ref{fig:5}a shows the spectral function $A(\vec k,\omega )=Im(G(\vec k,
\omega ))/\pi$ for $x=0$ at $\vec k=(\pi /2,\pi /2)$ at a series of 
temperatures.  The spectrum is split into upper and lower Hubbard bands, with a 
gap approximately $2\bar\Delta$.  The short-range order gap $\bar\Delta$ is
plotted in Fig.~\ref{fig:5}b; it is finite for all temperatures, but increases
significantly close to the mean-field N\'eel temperature (arrows).  The net
dispersion for two dopings, $x$ = 0, -0.15, is shown in Fig.~\ref{fig:6}; it is
in good agreement with the experimental\cite{nparm} and mean-field\cite{KLBM}
results.  The build up of hot spot magnons is reflected in the growth of
$Im(\Sigma (\vec k,\omega ))$ near $\omega =\xi_{\vec k+\vec Q}$, 
Fig.~\ref{fig:5}c (note the logarithmic scale).

\section{Implications for Hole Doping}

One expects, and observes, a certain degree of symmetry between electron and
hole doping: there is a susceptibility plateau associated with hot spots, 
Fig.~\ref{fig:1}, which terminates near optimal doping; the termination of 
magnetic order in electron-doped cuprates at a QCP near optimal doping is
matched in hole-doped cuprates by the observation of a QCP near optimal doping
which appears to be associated with loss of magnetic correlations\cite{Tal1};
and in both cases an optimal, probably d-wave superconductivity is found near 
the QCP.  Also, it is expected that $U$ would have a similar decrease with
doping for either electron or hole doping (see the model calculation, dashed
line in Fig.~\ref{fig:1}).

On the other hand, there are also significant differences, not the least of
which is the {\it magnitude} of the superconducting $T_c$.  Most significantly,
there is considerable evidence for the appearence of nanoscale phase separation
-- either in the form of stripes\cite{Tran} or blobs\cite{Dav} -- for hole doped
cuprates, while the evidence is weaker or absent for electron 
doping\cite{AIP,AKIS}.  As noted above, this difference arises naturally
within the present calculations, which find instability of the commensurate
magnetic order for hole doping only (e.g., Fig.~\ref{fig:3}).

Given these similarities and differences, the present calculations can shed
some light on aspects of hole-doped physics:

(1) A large {\it pseudogap} is present at half filling, associated with 
short-range magnetic order.  Just as for electron doping, it should persist
with hole doping until short-range magnetic order is lost.  The 
observed\cite{Tal1} connection of the loss of magnetic fluctuations with the 
collapse of the pseudogap near $x=0.19$ strongly suggests an identification of 
the observed pseudogap at $T^*$ with the Mott pseudogap, as found for electron 
doping.  While a number of theories have proposed that the pseudogap is
associated with superconducting fluctuations, these seem to turn on at a
temperature lower than $T^*$.\cite{strcom}.  Remarkably, despite the 
complications associated with stripes, the mean-field transition temperature is 
within a factor of two of the observed pseudogap temperatures, 
Fig.~\ref{fig:2}b.

(2) The present calculations point to a close connection between the Van Hove
singularity (VHS) and the instability of commensurate magnetic order,
Fig.~\ref{fig:1}b.  This strongly suggests that the VHS is responsible for
the asymmetry between electron and hole doping, and in particular for any
frustrated phase separation.

(3) If the stripes are associated with frustrated phase separation, it is
important to identify the second (metallic) phase and understand what 
interaction stabilizes it (particularly since this is likely to be the phase
in which high-$T_c$ superconductivity arises).  In a purely Hubbard model,
this would be a ferromagnetic phase, hence probably incompatible with
superconductivity.  However, the reduction of $U$ with doping found here
strengthens the case for a nonmagnetic charge stripe associated with
interactions beyond the Hubbard model\cite{berma}.

\section{Discussion: Polaron Limit}

In the very low doping limit, isolated charge carriers should form (spin or
charge) polaronic states for either sign of doping.  The asymmetry between
hole and electron doping would then be reflected in interpolaronic interactions
being strongly attractive for hole doping.  For electron doping, the isolated
polarons could be much more easily pinned in the AFM background, leading to the
much stronger localization found in NCCO\cite{loccit}.

\section*{Acknowledgments}

This work has been supported in part by the Spanish Secretaria de
Estado de Educaci\'on y Universidades under contract n$^o$ SAB2000-0034.  The
work was carried out while I was on sabbatical at the Instituto de Ciencia de 
Materiales de Madrid (ICMM) in Madrid, and the Laboratory for Advanced Materials
at Stanford.  I thank Paco Guinea, Maria Vozmediano, and Z.X. Shen for inviting
me, and for numerous discussions.

\section*{Appendix: c-Axis Coupling}

\renewcommand{\theequation}{A.\arabic{equation}}
\setcounter{equation}{0}

A toy model is introduced to study the effect of interlayer coupling on
generating a finite N\'eel transition temperature $T_N$.  The interlayer hopping
is assumed to be a constant $t_z$ independent of in-plane momentum.  While a
term of the form $t_z(c_x-c_y)^2$ would not greatly change the results, in the
physical cuprates alternate CuO$_2$ planes tend to be {\it staggered}, which
should lead to frustration $t_z(\vec Q)=0$, and greatly reduced interlayer
coupling.  Indeed, in the cuprates it is entirely possible that interlayer
coupling is negligible, and that the N\'eel transition is actually of
Kosterlitz-Thouless form, due to reduced spin dimensionality caused by
spin-orbit coupling\cite{Ding}.  

Nevertheless, it is instructive to see how constant-$t_z$ interlayer coupling
can generate a finite $T_N$.  The revised Eq.~\ref{eq:4} can be written in the
symbolic form
\begin{equation}
Z\delta =\bar\delta_0+{T\over T_0}ln({D_0\over D+2\delta}),
\label{eq:4b}
\end{equation}
where $T_0=\pi^2A/6ua^2$ and $D_0=4CT/e$ are (doping-dependent) constants 
(Eq.~\ref{eq:4} -- the extra $\pi /2$ in $T_0$ coming from the $q_z$-integral) 
and\cite{STeW} $D\propto t_z^2$.  Thus for finite $t_z$, there is a non-zero
$T_N$ given by the solution of $\bar\delta_0+T/T^*_0=0$, with $T^*_0=T_0/\ln
(D_0/D)$.  For the calculation in Fig.~\ref{fig:2}a, a constant $T^*_0=1200K$
was assumed, but it is interesting to note that when the correct doping 
dependence of the parameters is included, $T_N\rightarrow 0$ as $A\rightarrow 
0$, suggesting that the much steeper falloff of $T_N$ with hole doping is
related to phase separation.  

Equation~\ref{eq:4b} can be rewritten using Eq.~\ref{eq:7}.  The N\'eel 
transition occurs when
\begin{equation}
J_z[\xi (T_N)/\xi_0(T_N)]^2=\Gamma T_N,
\label{eq:22a}
\end{equation}
with $J_z/J=(t_z/t)^2$, $\Gamma =16C/edU$, $d=D/t_z^2$ and $J=4t^2/U$,
suggestive of a form of interlayer coupling proposed earlier\cite{BiGGS}.

\end{document}